# First ground-level laboratory test of the two-phase xenon emission detector RED-100

**D.Yu. Akimov[a,b*], V.A. Belov[a,b], A.I. Bolozdynya[a], Yu.V. Efremenko[f], A.V. Etenko[a,d], A.V. Galavanov[a,c], D.V. Gouss[a], Yu.V. Gusakov[a,c], Dj.Ed. Kdib[a], A.V. Khromov[a], A.M. Konovalov[a,b,e], V.N. Kornoukhov[a,g], A.G. Kovalenko[a,b], E.S. Kozlova[a,b], A.V. Kumpan[a], A.V. Lukyashin[a,b], Yu.A. Melikyan[a,g], V.V. Moramzin[a], O.E. Razuvaeva[a,b], D.G. Rudik[a,b], A.V. Shakirov[a], G.E. Simakov[a,b,e], V.V. Sosnovtsev[a], Yu.V. Stogov[a], A.A. Vasin[a]**
**The RED-100 Collaboration**

[a] *National Research Nuclear University "MEPhI" (Moscow Engineering Physics Institute), Moscow, 115409, Russia*
[b] *Institute for Theoretical and Experimental Physics named by A.I. Alikhanov of National Research Center "Kurchatov Institute", Moscow, 117218, Russia*
[c] *Joint Institute for Nuclear Research, Dubna, Moscow region, 141980, Russia*
[d] *National Research Center "Kurchatov Institute", Moscow, 123098, Russia*
[e] *Moscow Institute of Physics and Technology (State University), Moscow, 117303, Russia*
[f] *University of Tennessee, Knoxville, TN 37996-1200, USA*
[g] *Institute for Nuclear Research, Moscow, 117312, Russia*
*E-mail:* akimov_d@itep.ru

ABSTRACT: RED-100 is a two-phase detector for study of coherent elastic scattering of reactor electron antineutrinos off xenon atomic nuclei. The detector contains a total of 200 kg of liquid xenon in a titanium cryostat with 160 kg of xenon in active volume inside a Teflon-made light collection cage associated with electrode system. The active volume is viewed by two arrays of nineteen 3"-diameter Hamamatsu R11410-20 PMTs assembled in two planes on top and bottom. The electrode system is equipped with an electron shutter (a patented device) to reduce a "spontaneous" single-electron noise. The detector was tested in a ground-level laboratory. The obtained results demonstrate that detection of coherent elastic scattering of reactor antineutrinos off xenon nuclei at Kalinin nuclear power plant with the RED-100 detector is feasible with a threshold of 4 ionization electrons.

* Corresponding author

## 1. Contents

## 1. Introduction

Recently observed process of coherent elastic neutrino-nucleus scattering (CEνNS) [1] was predicted in the Standard Model more than 40 years ago [2]. It was pointed out that a neutrino interacts coherently via exchange of Z-boson with all nucleons in a nucleus. The cross section of this process is approximately given by the formula [3]:

$$\sigma \approx 0.4 \cdot 10^{-44} N^2 (E_\nu)^2 \text{ cm}^2,$$

where $N$ is the number of neutrons in a nucleus and $E_\nu$ is the neutrino energy (in MeV). The formula is valid for the neutrino energies of up ~ 50 MeV. Due to the $N^2$ dependence this process has very large cross-section in compare with other known neutrino interactions.

CEνNS plays the dominant role in the Universe in the processes in which the intense neutrino fluxes are involved. Supernova bursts, for example, release ~ 99% of their energy to the neutrino radiation. Thus, precise experimental measurement of the cross-section of this process is very important for astrophysics. Study of the CEνNS process is considered not only as a confirmation of the Standard Model, but also as a probe for new physics beyond it. Another stimulating factor that drives the studies of the CEνNS process is a potential possibility to build detectors of a new generation on its basis for nuclear reactor monitoring purposes.

There was no experimental evidence for this process until recently because of technical difficulties in its detection: the energy transferred from the neutrino of a MeV-energy to the recoil nucleus is of an order of hundreds eV. To observe such nuclear recoils, a detector with a mass of more than several kilograms having a sub-keV energy threshold and working in a low-background environment is required which is rather challenging. Only in 2017, the process was observed by COHERENT international collaboration [1] at an intense pion decay-at-rest source (with neutrino energies up to 50 MeV) of the SNS accelerator facility, Oak Ridge, USA.

A worldwide experimental program for study CEνNS is under way [4], [5], [6], [7], [8], [9], [10], [11]. See, also, a review of these experiments in [12]. In this study, noble liquid gas detectors should play a decisive role similarly as happened in the search for WIMPs (Weakly Interacting Massive Particles). The techniques of WIMP and CEνNS detection are very close to each other because selection of only nuclear recoil events is required in both cases. Starting from 2005, approximately, the rate of setting experimental limits on the cross-section of spin-independent WIMP-proton interaction has rapidly increased namely due to the use of two-phase (or dual-phase) liquid xenon (LXe) emission detectors [13]. The main reason of this, first of all, is that such detectors can be easily scaled up in mass without sectioning. Combined with a 3-D coordinate sensitivity this allowed to select a fiducial volume inside a detector which is well shielded actively from the outer radioactive background (so-called "wall-less detector" conception). The second reason is the capability of noble gas detectors to identify the different kinds of particles in order to suppress the background of gamma-rays or electrons. Nowadays, detectors of a many-ton-scale are at a development or a construction stage [14], [15].

A technology of noble gas two-phase emission detector was proposed for the first time at MEPhI about 50 years ago [16]. In such a detector, a working substance (detection medium or target) is a condensed noble gas where the ionization electrons are produced, while a gas phase above the surface of liquid to which they are pulled out from the liquid by applied electric field, serves for detection of these ionization electrons [17], [18]. This is because of well developed methods that can be used for track detection in a gas phase. Such a detection conception is a liquid noble gas time projection chamber (TPC) which differs from a traditional gas TPC by the use of a liquid working medium inside an electric field-shaping cage. The coordinate along the charge drift (vertical) is obtained by taking into account a time stamp of interaction ("start") given by a scintillation signal. A "stop" time stamp and X&Y coordinates are provided by a position sensitive detector of ionization in a gas phase. Traditionally in two-phase emission detectors, an electroluminescent (or proportional scintillation) process with detection of the produced VUV light by a PMT array is used for this.

The RED-100 two-phase emission detector has been built at NRNU MEPhI by the RED Collaboration of three Russian institutions: NRNU MEPhI, NRC "Kurchatov institute" – ITEP and NRC "Kurchatov institute". It is the first two-phase emission detector in the world that can be used in conditions close to those of a ground-level laboratory (see description of the detector in [19], [20], [21]). The main goal of experiment is to study coherent scattering of electron antineutrino off xenon atomic nucleus in a close vicinity to an industrial nuclear reactor core at the Kalinin nuclear power plant (KNPP).

In this paper, we describe the first laboratory tests of the RED-100 detector in a ground-level laboratory and preliminary results of data analysis.

## 2. The RED-100 setup

The RED-100 experimental setup (figure 1) includes the following units and systems:

the RED-100 detector unit (4); the interface unit (5); the gas storage (7,8) and purification system (6); the cryogenic system: thermosyphon (2) and its control unit (9); the data acquisition system (10).

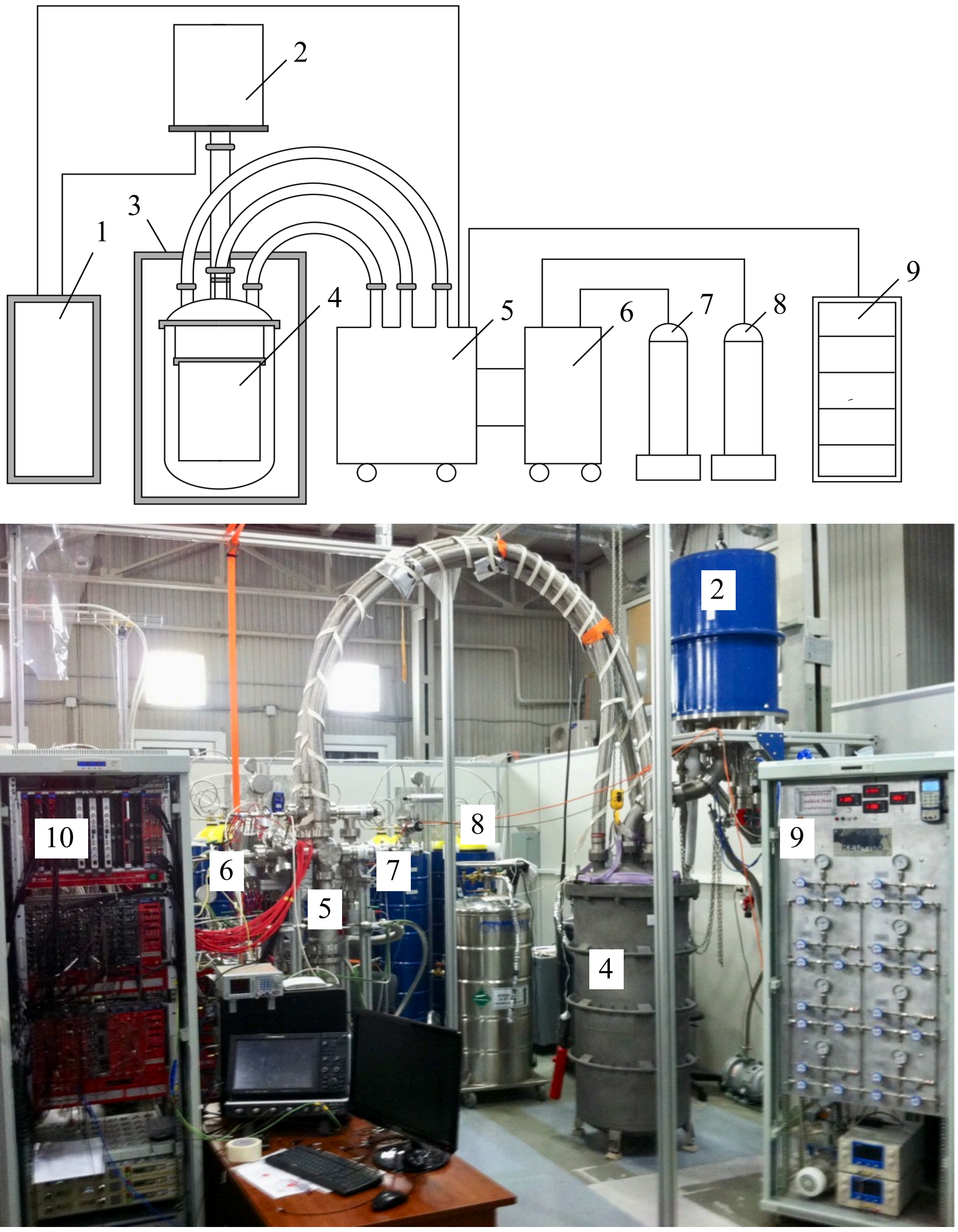

**Figure 1.** Block diagram (on top) and general view (on bottom) of the RED-100 setup: 1 – liquid nitrogen Dewar, 2 – cryostat of the thermosyphon cryogenic system, 3 – passive shielding, 4 – detector RED-100, 5 – interface unit, 6 – gas purification system, 7–8 – storage system for xenon gas, 9 – thermosyphon control system, 10 – data acquisition system.

One can find description of several setup elements in our previous publications [22], [23], [24], [25]. Here we give a description of the elements that were omitted earlier or given in insufficient details, as well as a description of the improvement of them.

## 2.1 The two-phase emission detector RED-100

The RED Collaboration has developed appropriate technical solutions applied to the RED-100 detector in order to meet specific requirements of operation at a site with a moderate (~ 50 m.

w. e.) protection from cosmic radiation. Such thickness of shielding (on average) is expected at KNPP. These are the following requirements:

- it is necessary to avoid illumination of photomultiplier tubes by intensive electroluminescent signals produced by energetic cosmic muons;
- when searching for the events with extremely low ionization, it is necessary to take special care for reduction of intensity of a single-electron (SE) noise caused by delayed emission of ionization electrons from the energetic events (mostly muons).

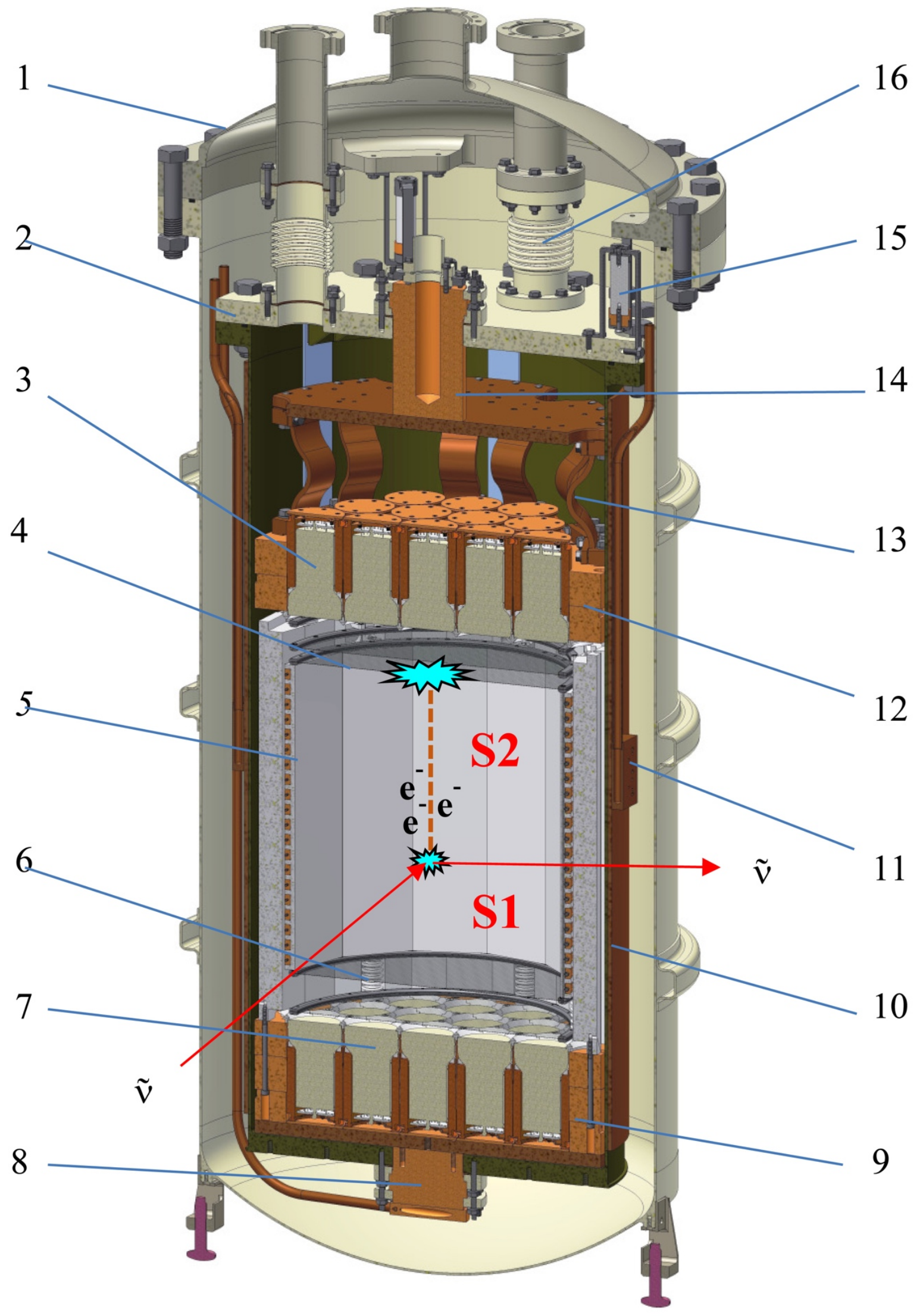

**Figure 2.** Schematic view of the RED-100 detector: 1 – external vessel of the cryostat, 2 – internal vessel of the cryostat, 3 – top array of 19 Hamamatsu R11410-20 photomultipliers, 4 – gridded anode and electron shutter, 5 – drift cage with Teflon reflecting walls, 6 – gridded cathode, 7 – bottom array of 19 Hamamatsu R11410-20 photomultipliers, 8 – cold head of the bottom thermosyphon, 9 – copper housing of the bottom PMT array, 10 – Copper screen of the internal vessel of the cryostat, 11 – cold head of the side thermosyphon, 12 – copper housing of the top PMT array, 13 – flexible heat bridge, 14 – top cold head for xenon condensation, 15 – Vespel made stand supporting cold vessel inside the external vessel of the cryostat, 16 – connection for cable channel; S1 – scintillation flash, S2 – electroluminescent flash.

The two-phase emission detector RED-100 (figure 2) contains ~200 kg of liquid xenon in a Teflon-made light collection dodecagon prism placed inside an electrode drift system. The liquid xenon TPC is viewed by two arrays of 19 Hamamatsu R11410-20 PMTs [26], [27], [28] and is installed into a titanium cryostat [29]. Detailed description of the RED-100 detector is given in [19]. Here we present an important patented improvement [30] of the electrode system which allowed us to reduce significantly the component of the SE noise produced by a spontaneous emission of the electrons trapped under the liquid surface (see [31] for details). The idea is to block collection of the ionization electrons from the drift volume when the events with the very high energy deposition take place. This is implemented by installing an additional grid (G1 in figure 3), a so-called electron shutter, in 3 mm below the extraction grid (G2). All grids in the detector are flat meshes with 4-mm-size hexagonal cells (optical transparence ~ 0.8) electrolytically formed from a stainless steel 0.2 mm foil. The meshes are spot-welded to stainless steel rings. In normal operation, the G1 grid has the same potential as G2 or is biased slightly negatively with respect to it, and the shutter is transparent to all electrons coming from the drift volume (see the results of simplified modelling of the electric field for wire grids in figure 4a). When the energy deposition exceeds a certain value, a trigger system based on scintillation signal produces a pulse to a special NIM unit which generates a 300-volt positive pulse applied to G1 through a capacitor. The shutter becomes closed (figure 4b) for the period equal to a duration of the pulse (typically 1 millisecond).

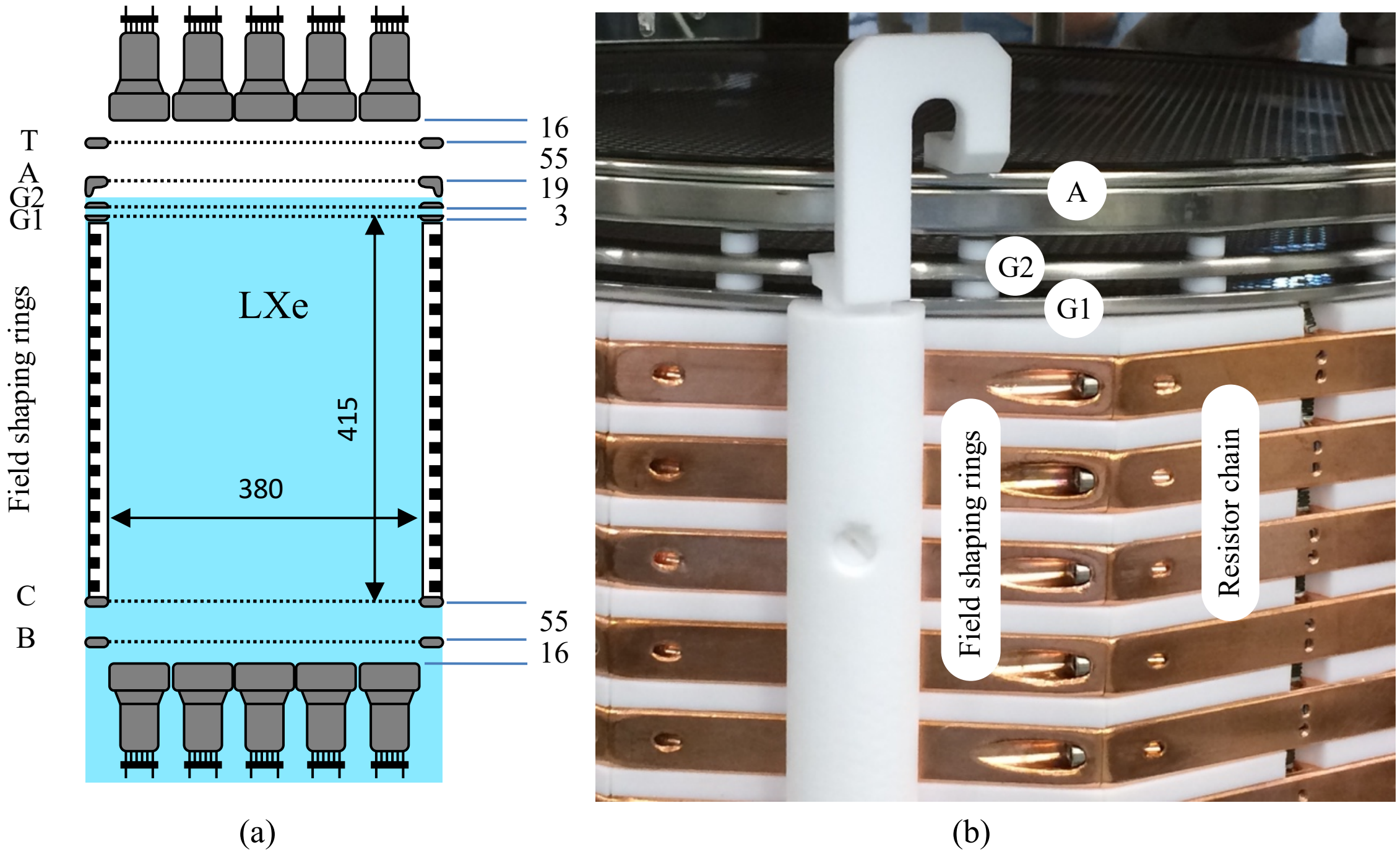

**Figure 3.** Schematic layout of grids (a) and photo of the top part of the electrode system (b): T and B – top and bottom grounded grids, A – anode grid, G1 – electron shutter grid, G2 – extraction grid, C – cathode grid; sizes of the drift volume and distances between grids are shown in mm.

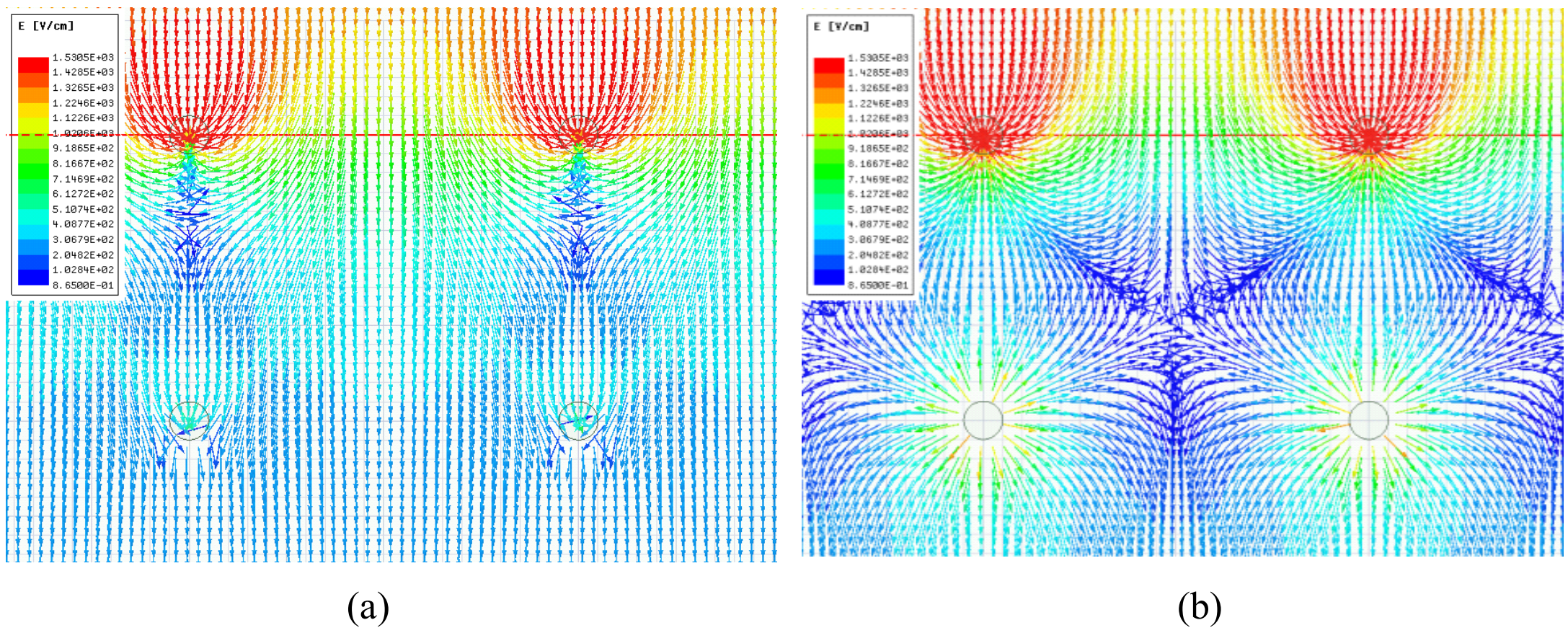

**Figure 4.** Simplified modelling of electric field in a wire electron shutter with a pitch of 4 mm and a distance between wire planes 3 mm using the ANSYS Maxwell® software package: (a) the shutter is open ($U_{G2} – U_{G1}$ = 50 V), (b) the shutter is closed ($U_{G2} – U_{G1}$ = 250 V).

The electron shutter also serves to satisfy the 1-st requirement: the illumination of PMTs by electroluminescence from muons is significantly reduced, because only the above-G2 part of the muon track produces the light. To avoid detection of this part of the light by PMTs, our previously developed method of PMT disabling by applying a positive pulsed bias (the same 300-V pulse) to photocathodes is used (see details in [26], [28]).

Another important improvement of the electrode system design is that the outer part of the anode ring is made thicker (see the scheme and the photo in figure 3) in order to be immersed into the liquid xenon. This prevents accumulation of the under-surface electrons between the rings of the A and G2 electrodes. The liquid xenon level is maintained at ~ 10 mm above the G2 electrode, the electroluminescent gap size is ~ 9 mm.

### 2.2 The xenon storage and purification system

For normal operation of the detector, xenon must be purified to the very high degree of purity. According to a worldwide practice, such very high purity is achieved by the use of continuous circulation of the working medium through a purifier in a gas phase. The RED-100 gas storage and purification system was designed with the use of this principle. It consists of all-metal gas and vacuum components: high-pressure gas cylinders for xenon storage placed on a weighing-machine platform, gas-supply and vacuum pipe-lines and valves, pressure and vacuum gauges, flowmeters, oil-free vacuum pumps, circulation pump, SAES MonoTorr (model PS4-MT15-R-2) hot metal getter.

### 2.3 The cryogenic system

In order to cooldown and maintain the cold vessel of the RED-100 detector at the liquid xenon temperature, a cryogenic system based on a thermosyphon (gravity assisted heat pipe) technology was built [22], [23]. This technology has several advantages in compare with other well-known thermostabilization systems: the reduced number of radioactive elements near detector active volume, the significantly low level of mechanical vibrations, the low refrigerant consumption and the ability to work without electric power. A thermosyphon of the RED-100 cryogenic system is a vertically oriented closed pipe which can be functionally divided into the following three parts: on top, a condenser immersed in a Dewar with free-boiling liquid nitrogen; on bottom, an evaporator (cold head) thermally contacted to the detector and a passive adiabatic part between these two parts. The pipe is filled with gaseous nitrogen. Effective heat removal from the detector is achieved by phase transition of nitrogen inside the thermosyphon. The gas condenses in the condenser and flows down to evaporator. There it boils up by the heat taken from the detector and rises upwards to the condenser where the process repeats. Thermal conductivity of thermosyphons can be very high, of an order of ~ several tens of kW/K·m, which is comparable to that of carbon nanotubes. Thermostabilization of the RED-100 detector is performed by varying a thermosyphon conductivity (cooling power) which, in its turn, is done by varying a mass of gaseous nitrogen inside it.

There are four thermosyphons in the RED-100 detector with a total cooling power of up to ~ 400 W. Cold heads of two of them (8 and 14 in figure 2) are installed in centers of the upper and lower flanges of the inner vessel of the cryostat. The top cold head (14) has a direct thermal contact with a copper disc inside the inner vessel for xenon gas condensation. This disc is also connected via copper thermal bridges (13) to the massive copper housing (12) of the top PMT array. The bottom cold head (8) is connected to another copper disc inside the inner vessel which is thermally connected to the bottom PMT array copper housing (9). There are two side cold heads (11) mounted on a copper screen surrounding the inner cryostat vessel in order to minimize temperature gradient in the radial direction.

The thermosyphon system can operate both in manual and automatic modes. A control rack of thermosyphon system (9 in figure 1) is equipped with stainless steel plumbing with manual and solenoid valves to supply nitrogen from a storage cylinder and release it from the thermosyphons to the atmosphere, and as well, to vacuum-pump the system. All manual valves are duplicated with Asco solenoid valves. The pressure inside the thermosyphons is measured by pressure transducers. Each thermosyphon operates independently. The electronic system for the operation of thermosyphons is based on ADAM-6000 series modules. These modules provide temperature measurements of the cold heads by PT-100 platinum sensors, measurements of nitrogen pressure inside thermosyphon tubes and driving of electromagnetic valves to inject or to discharge nitrogen into/from thermosyphons. All system parameters are automatically recorded in a database.

The thermosyphon cryogenic system allows to maintain the liquid xenon inside the RED-100 detector at a temperature ranged from 165 to 175 K with an accuracy of <0.2 K and to provide a vertical temperature gradient < 1 K/m. This allows to minimize convection flows of liquid xenon inside the sensitive volume that may affect to detector characteristics.

### 2.4 The electronics and the data acquisition system

The photomultipliers used in the RED-100 detector are Hamamatsu R11410-20 low-background devices designed to work in liquid xenon. PMT of this type has a metal body internally connected to the photocathode. In order to prevent possible electrical discharges inside the detector positive polarity HV is used. The PMTs are powered by a resistive voltage divider assembled on a printed circuit boards made of CIRLEX® polyimide material. SMD resistors, capacitors and receptacles for the PMT leads are mounted on the boards using a lead-free solder featuring minimal radioactivity. The Hamamatsu R11410-20 PMTs provide the high internal amplification factor ($\sim 10^6$). The single photoelectron (SPE) pulses have amplitude of ~ 1 mV and width of 10 ns.

To reduce cabling inside the detector volume, each PMT is connected with only one coaxial cable used both for signal read out and HV supply. The PMT cables are passed through a 4-m long flexible hose and are connected to a multipin vacuum feedthroughs at the interface unit (5 in figure 1). On the atmosphere side, the signals pass through so-called junction boxes which provide connection to outer electronics. Each box contains a decoupling capacitor, a passive protection circuit from possible HV breakdown and a controllable attenuator. There are six channels in each box and a test input for calibration of electronics. The PMT signals are amplified by fast low-noise 8-channel amplifiers Phillips Scientific 777 with bandwidth 0 – 200 MHz and adjustable gain from 2 to 50, having wideband noise < 25 μV RMS, referred to the input. Each amplifier has two outputs, one connected to the ADCs and another one connected to the linear Fan-out Phillips Scientific 748 unit, which is used to split signals for trigger circuits. The Flash ADC CAEN V1730 units are used for digitization of waveforms. These devices have sampling frequency 500 MHz with 14 bits per sample and allow to keep 5 million samples for each of 16 channels.

The RED-100 detector operates with a wide dynamic range of signals. This is required since the energy deposition for the detected events varies significantly. There are the following types of events to which the detector is sensitive:

- events in keV and sub-keV energy region, the region where the CEνNS signal from reactor antineutrinos is expected, the ionization signal in this region is of an order of just a few electrons;
- events with energies from several keV to several MeV; the ionization signal in this region is from few dozens to ~ $10^5$ electrons – region of gamma events;
- events with energies > 10 MeV (> $10^6$ electrons) from cosmic muons crossing the detector.

The signal from one electron (SE - single electron) is represented by a bunch of SPE pulses from different PMTs randomly uniformly distributed within ~ 2 μs interval that equals to a duration of electroluminescence in 9-mm gas gap. The total duration of the event waveform (300 μs) was chosen to cover the maximal drift time for ionization electrons in the sensitive volume which is about 250 μs. The choice of electronics for the RED-100 detector was governed by these requirements, with the main aim to distinguish SPEs over period of maximal drift time. The DAQ system records waveforms for each PMT on an event-by-event basis and store it to disk for offline analysis.

A trigger system of the RED-100 detector should operate both with a low threshold and in a wide dynamic range of signals. To fulfill these requirements the trigger is produced with the use of CAEN V1495 customizable FPGA unit [24]. The operation mode with the highest sensitivity is optimized for signals having from 1 to 10 ionization electrons. It is based on counting within a 3-μs window the output pulses from the CAEN V895 leading edge discriminators in each channel with a threshold of ≈1/3 of SPE amplitude. The top and bottom PMT arrays are counted separately. A trigger is formed if the number of counts lies within preset limits. This allows to avoid random coincidences and triggering from large signals.

For selection of gamma events, an analogue sum using CAEN N625 units is produced separately for the top and bottom PMT arrays. These signals are digitized by additional ADCs, and the waveforms are sent directly to the trigger unit. Online digital processing of waveforms allows to select the events with coincidences of scintillation and electroluminescence signals within the maximal drift time. Cosmic muons are selected by a coincidence of large scintillation signals in several PMTs in the bottom array. The trigger unit generates a gate pulse with necessary duration for switching off the PMTs and closing the electron shutter. Also, the unit monitors the detector performance by measuring the individual PMT SPE rates.

The analogue part of electronics is assembled in two NIM crates and resides in a standard 19" rack. The Flash ADC, trigger, and discriminator units are installed in a VME crate placed in the same rack and controlled by a computer. The waveforms and the information from the trigger unit are read out with the computer and stored to the hard drive. The High Voltage power suppliers for PMTs and drift electrodes are located in the same rack and controlled by a computer.

## 3. Preparation of the RED-100 setup for operation

A full-scale test of the RED-100 experimental setup (figure 2) was carried out in order to verify the developed technical solutions. The test included the following several stages:

1. Preparation of the working medium - 200 kg of xenon [32].
2. Preparation of the detector for filling up with the liquid xenon, including cleaning of the internal surfaces, testing and tuning photomultipliers.
3. Cooling down the detector and condensing xenon.
4. Engineering run which included physical measurements of the cosmic muons, the gamma-rays from radioactive sources $^{22}$Na and $^{60}$Co and measurements of the single-electron noise.
5. Continuous circulation of the working medium (transferred to a gas phase) through a chemical getter during all physical test.
6. Evacuation of xenon into high-pressure cylinders after completion of the engineering run.
7. Analysis of the results.

Preparation of the experiment (steps 1 and 2) took about 3 months. Cooling the detector and filling it with total amount of liquid xenon took 3 days. The timeline and operating modes of the setup are shown in table 1. Below, the most critical stages of preparation of the working medium and the RED-100 detector are described.

Table 1. The main modes of operation of the RED-100 during test and the main parameters.

| Mode | *T*, °C | Active period, hours | Active thermo-syphons | *dT/dt*, degrees/hour | Heater | Vacuum jacket | Pressure of Xe, atm |
|---|---|---|---|---|---|---|---|
| 1. Cooling down | +20 ÷ -103 | 20 | All | -5 | 0 | 0 | 1 - 1.3 |
| 2. Stabilization | -103 | 2 | All | 0 on average | 0 | 0 | 1.3 |
| 3. Xenon condensation | -103 | 60 | All, but mainly top | 0 on average | 0 | 0 | 1 - 1.3 |
| 4. Circulation of xenon | -103 | Continuously | All, but mainly top | 0 on average | 0 | 0 | 1.3 |
| 5. Removing LXe from the detector | -100 | 20 | 0 | 0 on average | on | 0 | 1 - 1.3 |
| 6. Warming up | -100 ÷ 0 | 20 | 0 | +5 | on | Nitrogen 1 atm | 0 |

### 3.1 Purification of xenon

The very high degree of xenon purity (at a level of ~1 ppb of electronegative and light-absorbing contaminations) is required in order to achieve the proper performance of the RED-100 detector. The xenon used in the detector is depleted in $^{136}$Xe: it has undergone a centrifuging isotope separation procedure. We have found that after passing the isotope separation procedure this gas became contaminated with a highly electronegative substance ($SF_6$ molecules and organic compounds of the $C_mH_nR_k$ class, according to the performed chemical analysis). The electron lifetime measured in the original gas sample was about ~ 0.1 μs (in liquid phase). This was explained by the presence of high-electronegative impurities presumably due to the use of a special fluorine-containing high-molecular-weight lubricant [33] in gas centrifuges. We have found that the SAES MonoTorr hot metal getter is ineffective for purification of such contaminated xenon gas. For purification of it, we used a spark-discharge technique described in detail in [34].

Initial purification of xenon was performed in two stages as described in [32]. At the first stage, the high-molecular-weight impurities had been decomposed under the irradiation of intensive ultraviolet light from an electric spark. At the second stage, a titanium nano-scale dust suspended in liquid xenon was used to absorb chemically active impurities. As a result, the original gas has been purified in total amount of ~ 200 kg to the purity level corresponded to > 100 μs of the electron lifetime in liquid xenon. This gas was used then to fill the RED-100 detector.

Further purification of xenon to the higher purity level than that achieved by a spark-discharge technique was performed by a continuous circulation of gas in a loop through the detector chamber and through the hot SAES MonoTorr getter [32]. Continuous purification allowed us not only to purify a medium but as well to clean up all the interior surfaces of the detector on which electronegative impurities may be absorbed.

The circulation of xenon was provided by a membrane pump KNF Neuberger N143SV.12E. The circulation purification was carried out in two stages. At the first stage, the detector at room temperature was filled with gaseous xenon at a pressure of 1.3 atm, and its internal surfaces were washed by gas circulating for about 1 month. At the second stage, after condensation of xenon into the detector circulation was performed with the use of liquid xenon. Xenon was taken from the detector as a liquid, then it was evaporated, then passed through the getter, and finally condensed back to the detector. A line with coaxial bellow hoses was used as a heat exchanger between two gas flows: the evaporated cold xenon and the coming to the detector one. The circulation flow rate was set at ~ 10 liters (at normal pressure) per minute, a maximal rate recommended for efficient purification of noble gases by the PS4-MT15-R-2 getter.

### 3.2 Cooling down the detector and filling with liquid xenon

Cooling down of the RED-100 detector was performed as follows. At the beginning, the thermosyphon Dewar was filled with liquid nitrogen. After that, all four preliminary evacuated thermosyphon pipes were filled by the nitrogen gas to a pressure of up to 4.5 atm. The cooling process started with an average cooling rate and ~ 15 K/h for the top and bottom thermosyphons and ~ 3.5 K/h for the side ones (see figure 5). When the temperature at the side thermosyphon cold heads reached T = -103°C (170 K) they were started to be maintained at this value. The main heat removal from the detector continued only by the top and bottom thermosyphons. After ~ 35 hours, the temperature of all cold heads reached the preset value $T$ = -103°C of stabilization that corresponds to the pressure $p$ = 1.3 atm of saturated xenon vapor. An interval of ± 0.2 K from the preset value was chosen for each thermosyphon in order to maintain this temperature. Thermostabilization by each thermosyphon operates as follows. When the temperature reaches the upper limit, an additional nitrogen is injected by opening a solenoid valve for 30 ms. If the temperature does not decrease within 30-40 minutes, the next injection of nitrogen is provided. This procedure is repeated until the temperature is stabilized in the specified interval. For the lower limit, the process is similar except the nitrogen is discharged from a pipe.

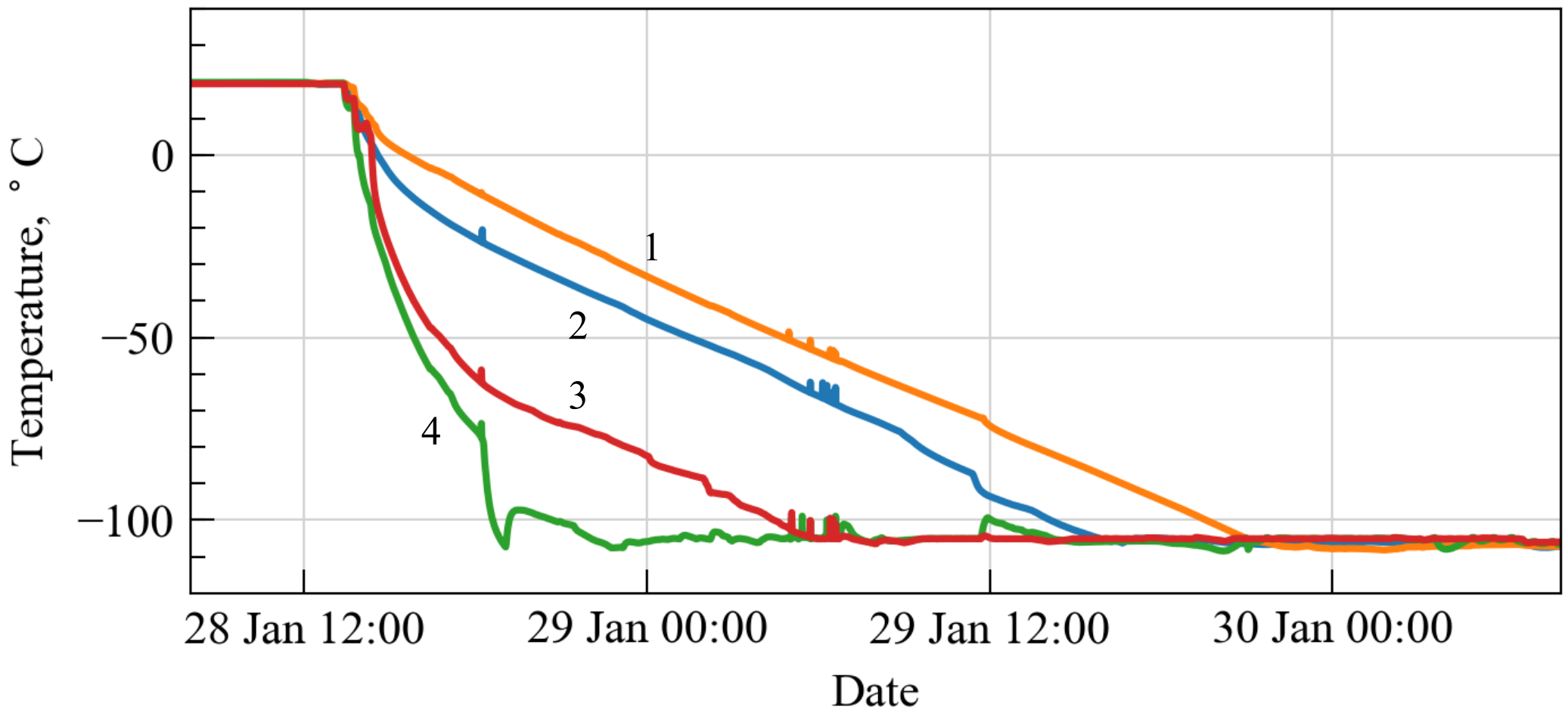

**Figure 5.** Temperature changes of thermosyphon cold heads during the cooling process; 1 – bottom, 2 – top, 3,4 – side.

After the temperature of cold heads was kept stable for ~ 5 hours at -103 °C xenon started to condense into the detector. The top thermosyphon was switched to the maximal cooling power at that time. This was done by increasing the nitrogen pressure. The inflow of xenon was performed through the SAES getter. The pressure of xenon in the detector was maintained in the range from 1.20 to 1.35 atm. The rate of xenon condensation was kept at ~ 3.3 kg/h (~ 10 slpm). The level of liquid xenon in the detector during condensation was continuously monitored by measuring three values: a mass of the residual gas in the cylinders, a capacitance of a cylindrical level meter installed in the detector, and the duration of muon signals produced in the above-liquid part of the drift volume (between G1 and C). When liquid xenon level approached the A-G2 gap (see figure 3), a precise measurement of the level was performed by obtaining the capacitance between the grid G2 and the anode A with the use of a charge sensitive preamplifier and a generator pulse. In addition, the distance between the liquid xenon level and the anode was estimated from a duration of electroluminescent signal for point-like events. Both measurements were in a good agreement. The process of filling had stopped when the level reached 9 mm above the G2 grid. This corresponded to ~ 200 kg of xenon contained in the detector. After that, the circulation process started with a flow rate of ~ 8.5 slpm. At this stage, the cooling power of the top thermosyphon was reduced because the heat exchanger operation started, and the detector stabilized at $T$= -103 ± 0.2°C and $p$ = 1.30 ± 0.02 atm which were maintained throughout the whole experimental run.

Detector levelling in horizontal plane was performed using comparison of the duration of electroluminescent signals from the corner photomultipliers of the PMT array hexagon.

### 3.3 Electron lifetime monitoring

Because of the availability of cosmic ray muons in great numbers we used them for the electron lifetime monitoring during the circulation process. It turned to be a very simple and convenient method. To simplify the recording and averaging of the RED-100 waveforms, we used the LeCroy WaveRunner 640Zi digital oscilloscope instead of data acquisition system. The PMT signals from the top and the bottom arrays were summed independently and digitized by two oscilloscope channels. The averaged waveforms were stored for further analysis. It can be shown that for our geometry of the sensitive volume and in the case of infinite electron lifetime the shape of the averaged waveform is rectangular for the uniform muon flux even if the individual averaged waveforms correspond to the tracks not completely fit between the G1 and C grids (muons travelling at large angles with respect to vertical). In the real case, when the capture of electrons by electronegative impurities takes place the averaged signal exponentially decays with time constant $\tau$ equal to the electron lifetime. This is shown by an example in figure 6b. The narrow peak at zero time is caused by triggering both from the scintillation and the short single electroluminescent signals (from point-like events). This example corresponds to $\tau$

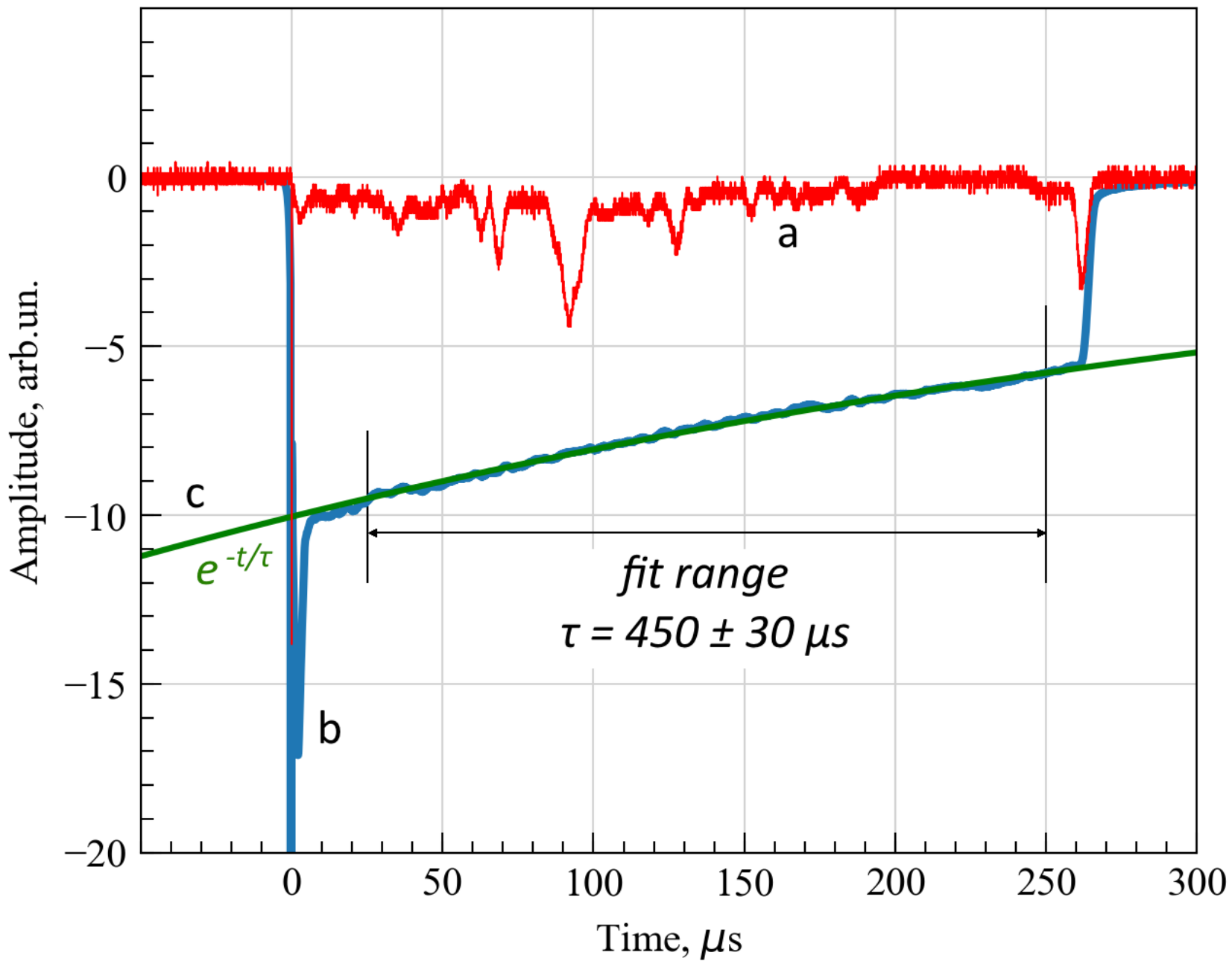

**Figure 6.** Examples of individual (a) and averaged (b) waveforms of muon events in the RED-100 detector, c – exponential fit of the averaged waveform; 10000 muon events were taken for averaging.

= 450 ± 30 μs which is greater than the total drift time between C and G1 (equal to 265 μs). A waveform for an individual event of the muon passed through both G1 and C grids is shows in figure 6a. Individual prongs of the waveform correspond to δ-electron blobs.

Evolution of the electron lifetime during the run is demonstrated in figure 7. One can see the practically stable increase of the lifetime from ~ 40 μs at the beginning to several milliseconds at the end of the run. Degradation of the lifetime at some periods was caused by technical reasons: adding some small portion of xenon on February 9, electrical discharge and adding another small portion of xenon on February 15, and replacement of broken membrane in the circulation pump on February 19.

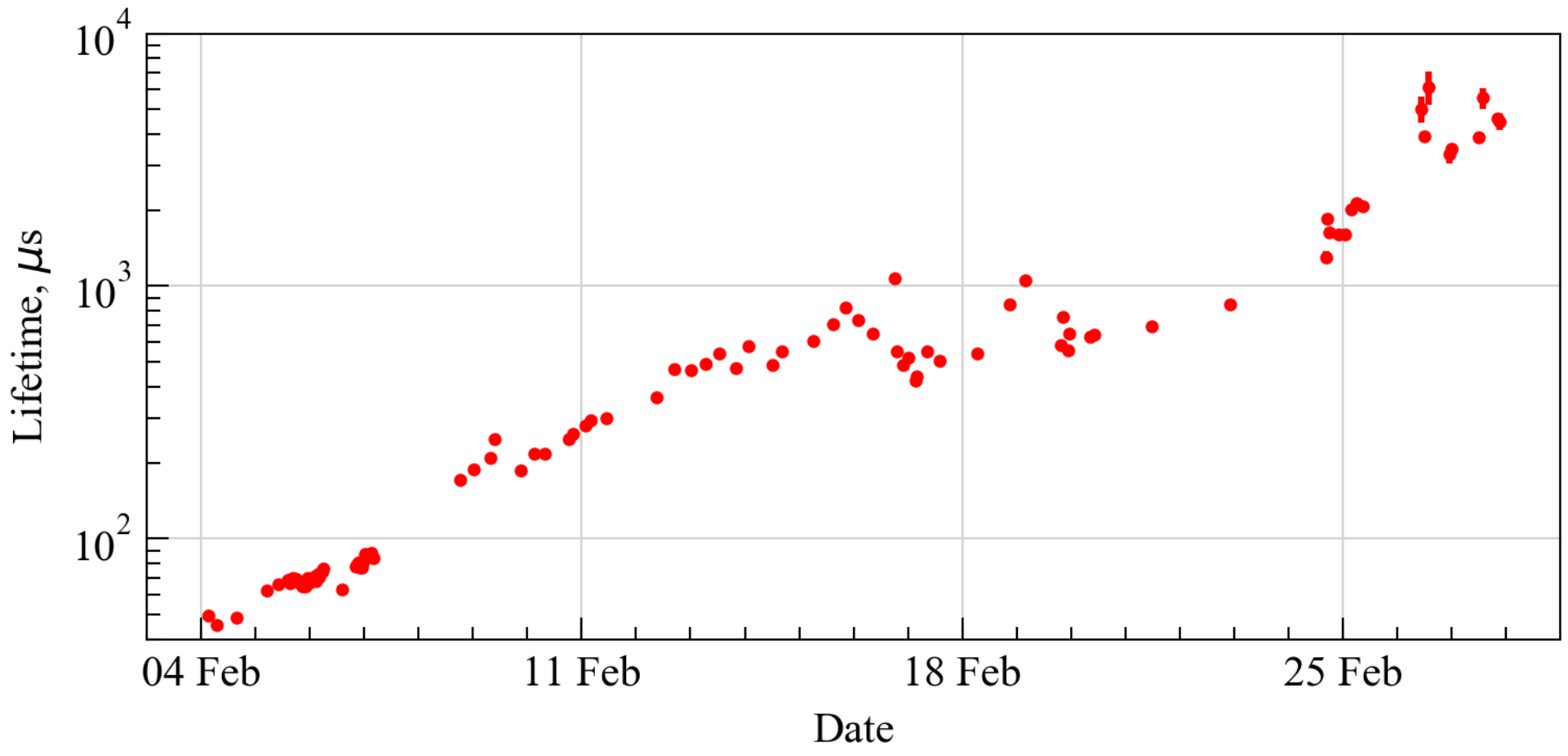

**Figure 7.** Evolution of the electron lifetime in the RED-100 detector during constant circulation of xenon in the period February 4 – 27, 2019. Electric field strength is 0.3kV/cm

## 4. Results of physical test

The main goal of physical test of the RED-100 detector was to demonstrate its sensitivity to the events typical for the CEνNS process, i.e. having energy deposition in a sub-keV energy range and composed only by few ionization electrons. To obtain the detector response in this energy region, one must perform the following necessary procedures:

- to obtain the energy scale of the S2 signals in photoelectrons (SPE) per keV,
- to identify single electron (SE) events and to measure them in terms of SPE number,
- to obtain experimentally all characteristics which are necessary for the full spatial and energy reconstruction of the events in all energy regions,
- to estimate the expected rate of background events in the CEνNS signal region at the experimental site.

To address these requirements, calibration runs with γ-sources and SE background taking runs were carried out.

### 4.1 Energy calibration by γ-sources

Energy calibration of the RED-100 detector was performed by $^{22}$Na and $^{60}$Co γ-sources. A collimated source was placed nearby the cryostat at a different height in the range between the levels corresponding to the G2 and C electrodes and at different locations around the cryostat. The trigger on γ-event was generated by the S1 signal crossing a preset amplitude threshold, and the additional requirement was that its duration to be less than 300 ns. A coincidence with a NaI[Tl] detector situated in opposite side from the source was used to select the 511-keV γ-events in case of $^{22}$Na source. The events were digitized in a 300-μs window with a trigger point set at 30 μs from the beginning of it.

The recorded events were analyzed with the use of a self-developed software package REDOffline. This tool was used for data reduction. In further analysis, subsequent clustering of the SPE pulses and classification of the obtained clusters as S1 or S2 signals according to their time structure were done. A typical S1 signal has a duration of ~ 100 ns, and S2 signal, of ~ 2 μs. Only the events started with S1 at the trigger point (and without S2 before trigger) were selected. The next selection criterion was the "proper" time interval between S1 and the S2 signal (or signals, in case of a multi-vertex event) corresponding to the vertical position of the γ-source. In the selected events, the S2 signals were corrected taking into account their vertical positions and the electron lifetime measured in the muon runs. For the $^{22}$Na source, only single-vertex events (mostly correspondent to a photopeak) were selected for simplicity.

The selected and corrected events were processed then with the use of an ANTS-2 software package [35] specially developed for event reconstruction in detectors of Anger camera type. The energy (in arbitrary units) of events and their coordinates in the XY (horizontal) plane were obtained at the output of this package. An example of XY coordinate reconstruction for gammas from the $^{22}$Na source is shown in figure 8.

There are very important characteristics obtained at the output of the γ-event reconstruction procedure called PMT light response functions (LRF). For each individual PMT, LRF is a relative dependence of the signal value versus the distance of a light flash from the PMT center in the XY plane. The shape of LRF depends on the PMT position in array. Example of LRF for one of PMTs from the top array is shown in figure 9. These functions were used for reconstruction of the SE events (see below).

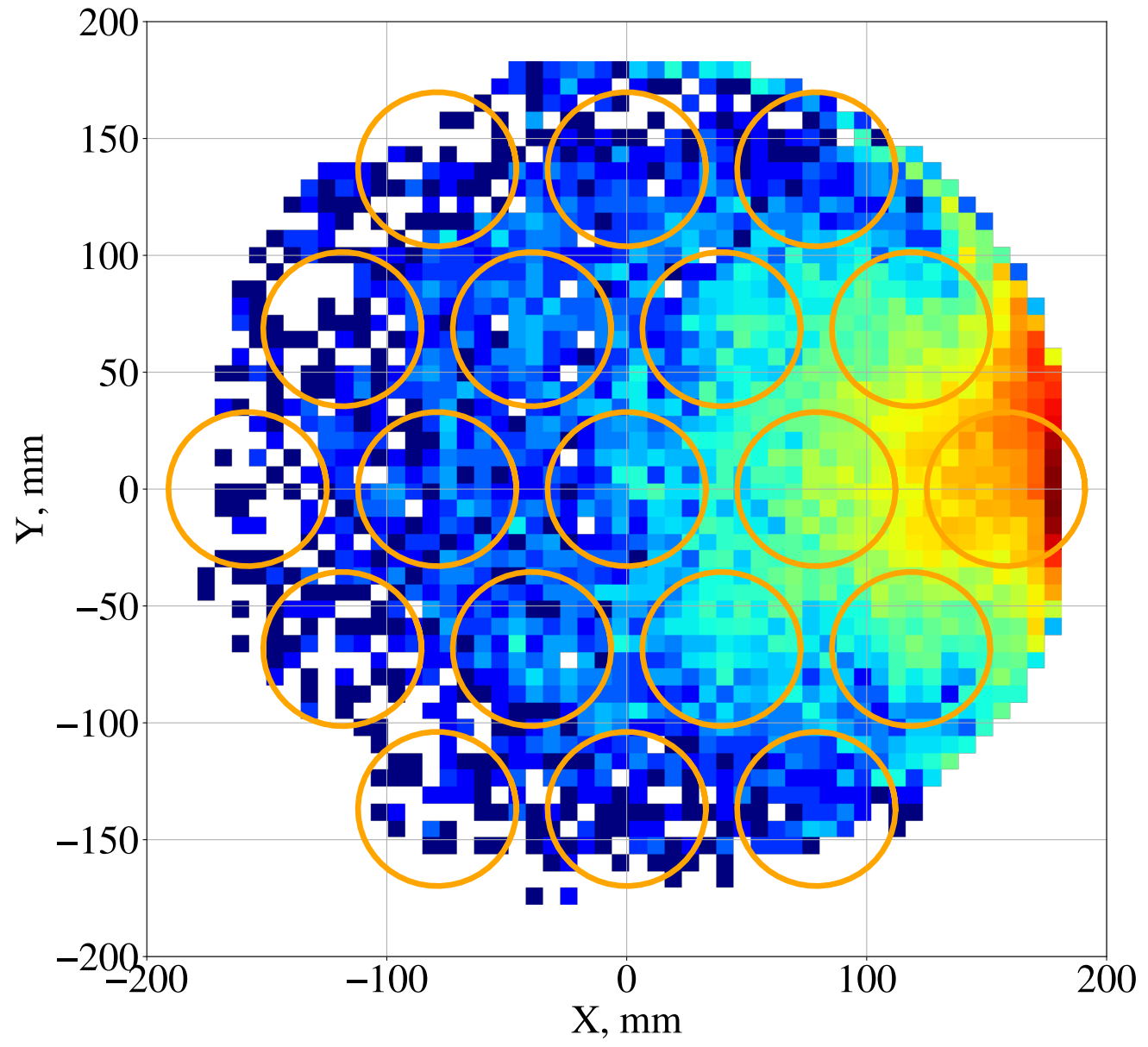

**Figure 8.** Distribution of γ-ray events from the $^{22}$Na source in the XY plane. The circles indicate the positions of the PMTs in the top array.

A value proportional to the energy released in ionization was obtained by ANTS-2 for each point of γ-ray interaction. Since the fractions of energy that go to scintillation and ionization are anticorrelated in a liquid noble gas media [36], [37] the new (reconstructed) S2 value is then

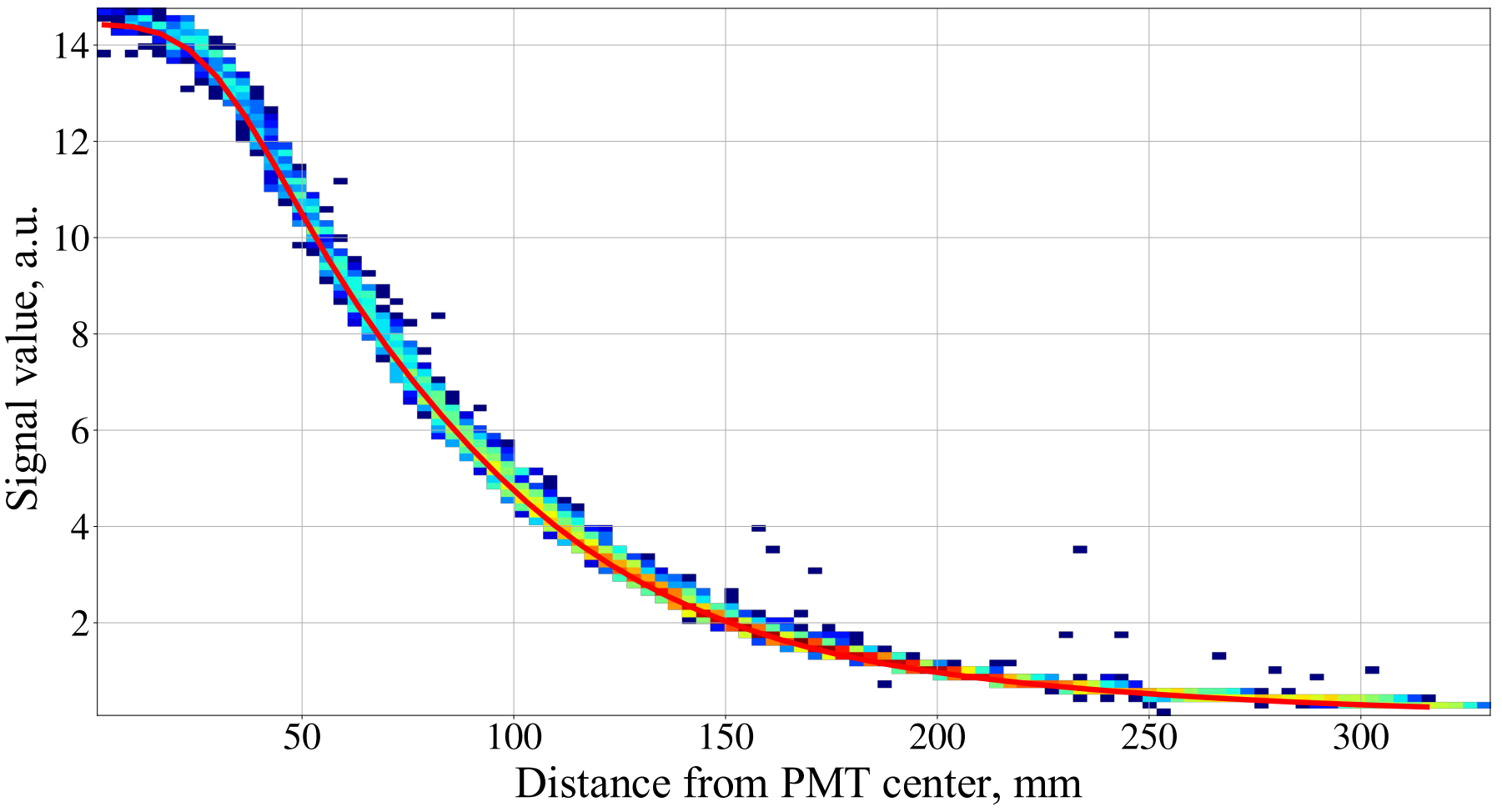

**Figure 9.** Example of LRF for one of PMTs of the top array located at the hexagon corner; red line is a spline fit.

linearly combined with S1 in order to obtain the value $A = a{\cdot}S1 + b{\cdot}S2$ proportional to a total energy. The coefficients $a$ and $b$ were adjusted so as to minimize a width of $A$ value distribution for the 0.511 MeV γ-line from the $^{22}$Na source (1 in figure 10). The same coefficients were used to produce the $A$ value for the $^{60}$Co lines (1.17 and 1.33 MeV; distribution 2 in figure 10). Note, that these lines cannot be resolved on separate S1 and S2 pulse area distributions. The obtained

energy resolution (FWHM) is 12.9 ± 1.2, 12.7 ± 0.47, and 11.28 ± 0.35 % for 0.511, 1.17, and 1.33 MeV, respectively.

A dependence of the obtained value $A$ on the energy of γ-rays is shown in figure 11. The good linearity of the response of the RED-100 detector was obtained.

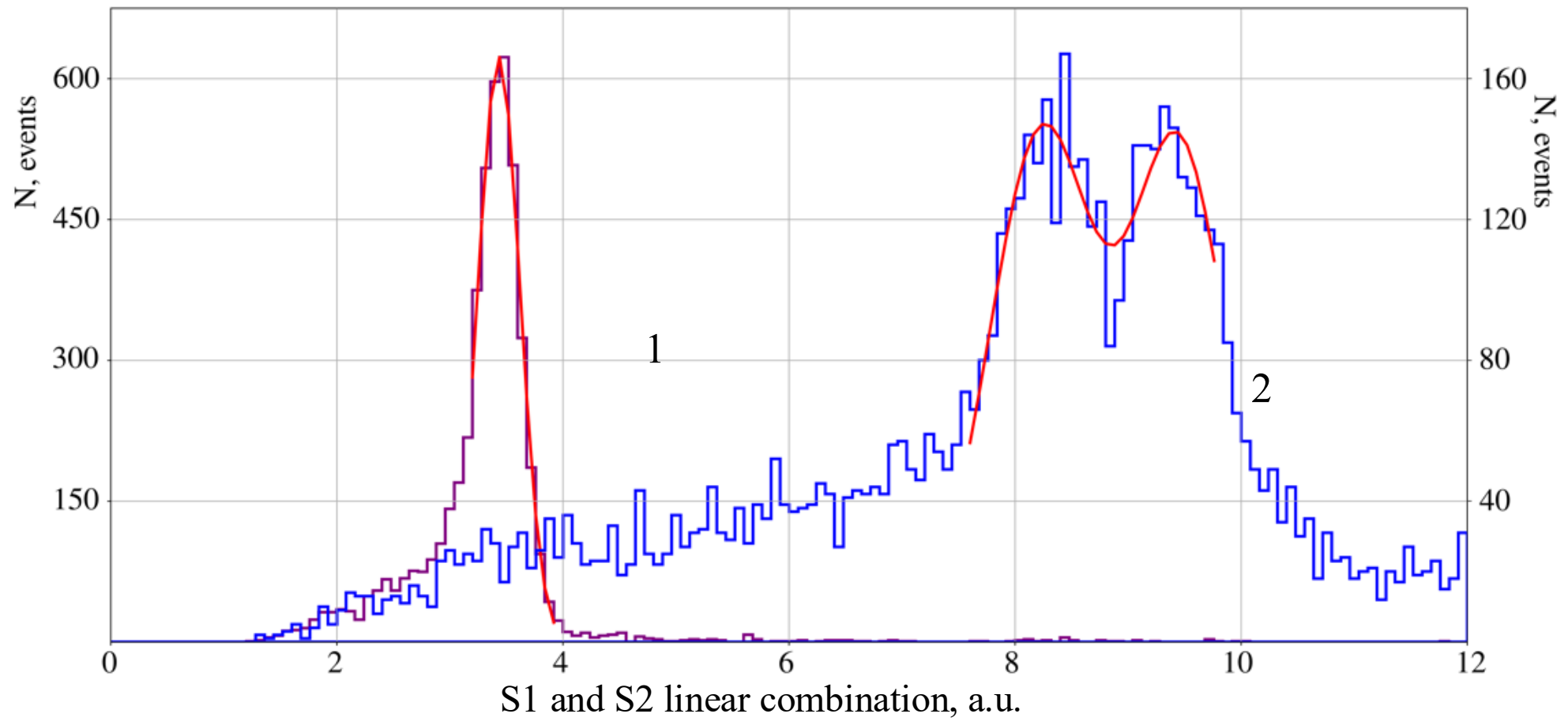

**Figure 10.** Distribution of combined S1 and S2 for $^{22}Na$ (1, left axis) and $^{60}Co$ (2, right axis) γ-sources.

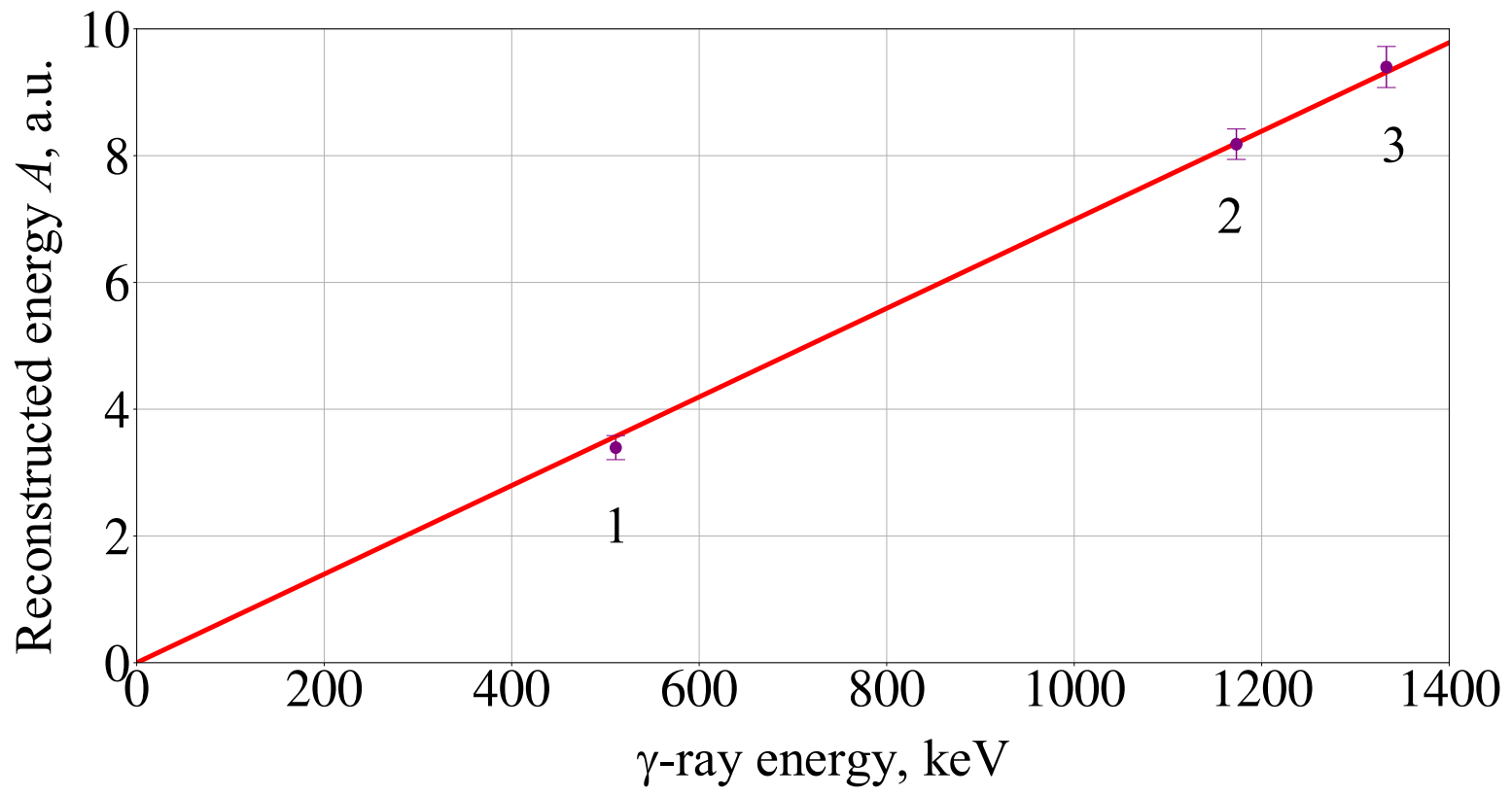

**Figure 11.** Reconstructed energy $A$ versus the energy of γ-rays; 1 – 0.511 MeV ($^{22}Na$), 2,3 – 1.17 and 1.33 MeV ($^{60}Co$).

### 4.2 SE calibration

The first generation of two-phase xenon detectors used for direct WIMP search and dedicated two-phase test chambers for laboratory R&D demonstrated the sensitivity of detectors of this type to single ionization electrons [31], [38], [39], [40], [41], [42]. Typical signal for a single ionization electron emitted from the liquid to the gas phase and drifting between the liquid surface and the anode is a sequence of SPE pulses with a characteristic duration of an S2 signal (~ 2 μs in our case). According to worldwide study, part of these SE originates from photo-ionization of detector materials and the impurities dissolved in xenon by the VUV light from S1 or S2 [38], [39], [40], [42]. The SE signals that appear without straightforward causal link to the presence of light in the detector are usually referred to as "spontaneous" ones [31], [39], [40], [41]. Their properties and hypotheses of their origin are discussed in ([18], page 24), [31], [39], [41], [43]. As pointed out in [43] and observed by us in this test with RED-100, there is a large excess of SE events after the energetic events, especially muons, with exponential decay of their intensity. In order to reduce the average rate of "spontaneous" SE, the readout electronics was blocked during 1 ms after the S1 scintillation signal from muons by applying the blocking pulse to the electron shutter. These tails originated from different events and mixed together may form this SE background. Regardless of the origin, these signals may be used for calibration of the detector response to the smallest ionization. This is of particular importance taking into account the fact that the signal from the coherent elastic scattering of reactor antineutrino off Xe atomic nucleus has a value of only a few ionization electrons [44].

In order to analyze single electron events, 300-μs waveforms with triggering on S1 signals from gammas were used. A trigger point was set at 25 μs before the end of waveform. Thus, pre-trace parts of waveforms were used as less contaminated by S1 or S2 pulses. These pre-trace parts of waveforms were examined to find SPE clusters correspondent to "spontaneous" SE signals having a characteristic duration of S2. An example of SE signal is shown in figure 12.

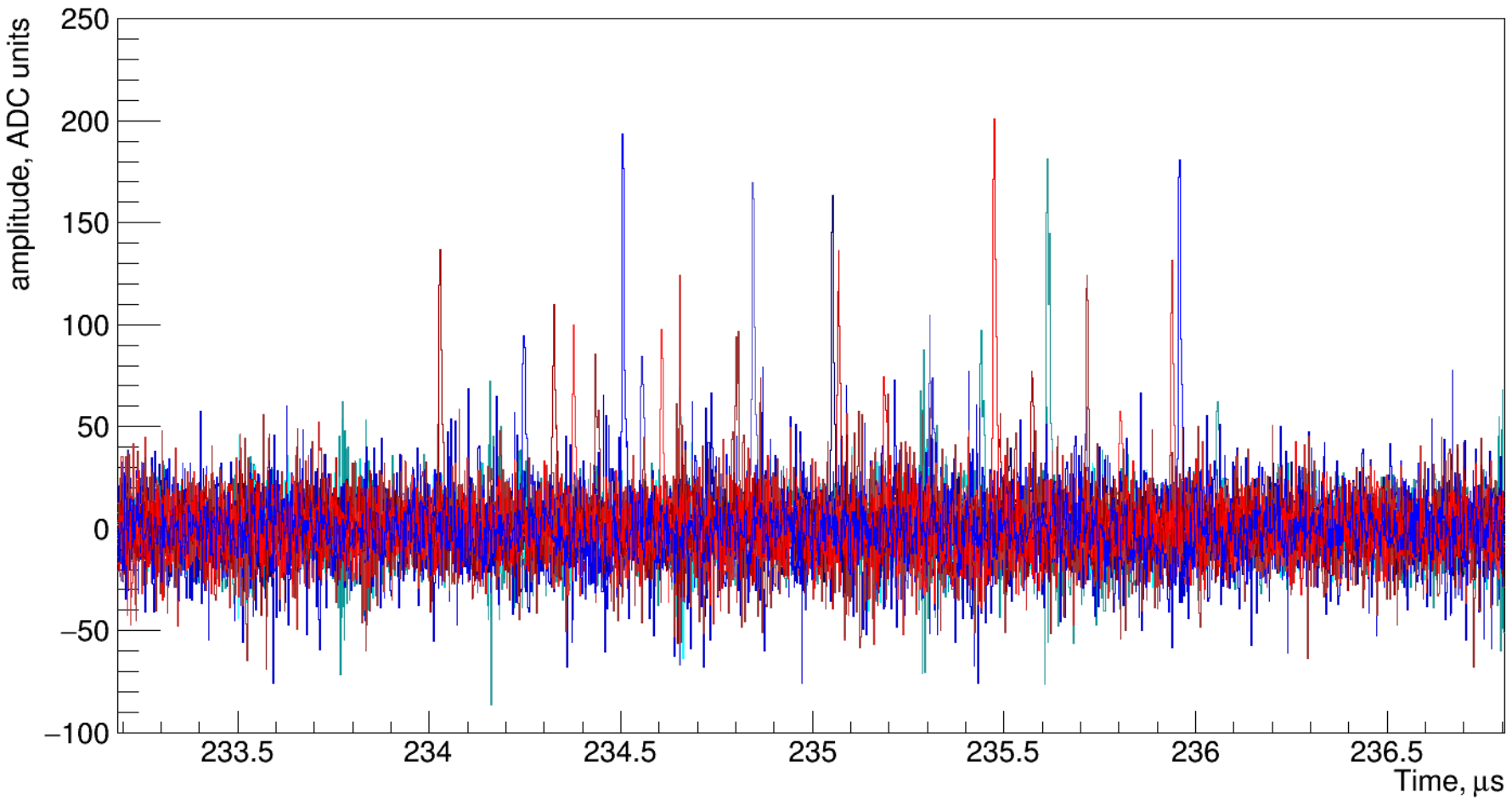

**Figure 12.** Example of SE signal (between 234 and 236 μs). The waveforms corresponding to different PMTs are overlaid.

The distribution of durations of the found SPE clusters is shown in figure 13a. One can see that the distribution peaks at 2 μs which is the same value as the duration of S2 obtained for the large signals. The right tail of distribution is explained by pile-ups of individual SE clusters because the count rate of them is rather high (see below). The clusters having duration from 1.4 to 2.6 μs (marked by vertical lines in figure 13a) were selected, and a total pulse area distribution was plotted for them (see figure 13b). The SE area (called also in literature as "a secondary-scintillation gain") of $29^{+6}_{-2}$ SPE was obtained from this distribution for extraction electric field strength of 3.0 ± 0.1 kV/cm in the liquid phase (5.55 ± 0.2 kV/cm in gas) and for the central region of detector restricted by a radius of 10 cm.

### 4.3 Electron emission efficiency

The efficiency of electron emission (EEE) from the liquid xenon surface to the gas phase under the applied electric field is a very important characteristic of a two-phase emission detector.

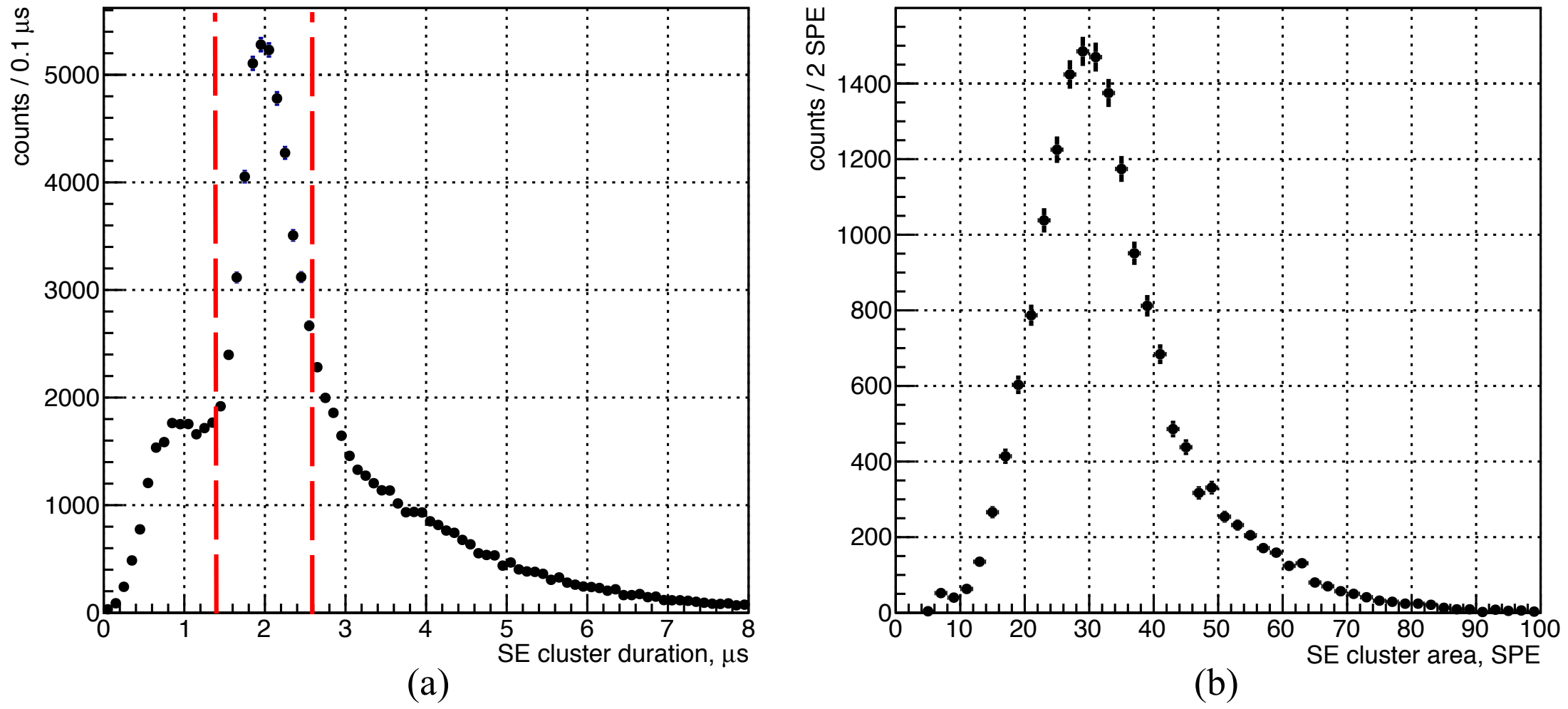

**Figure 13.** Distribution of durations of the SPE clusters (a) and distribution of areas of selected SPE clusters (b) at 3.0 kV/cm (in the liquid phase).

There are many experimental tests in which this value has been measured [42], [45], [46], [47], [48]. However, the results are quite different from each other. We obtained the EEE value from our own γ and SE calibration. This value equals to the number of electrons emitted to the gas phase divided by the number of electrons created in ionization process. The first number (for the 511-MeV γ-line) was obtained by dividing the peak position on the S2 area distribution expressed in SPE by the SE area expressed in the same units. To find the second one, we used prediction by NEST [49], [50] for electron recoils from gammas of this energy and for the electric field strength in the sensitive volume equal to our working value of 0.217 kV/cm. The obtained EEE value equals 0.54 ± 0.08. This value, however, is by a factor of ~ 1.5 higher than that obtained in [42], [45], [46] (at 3.0 kV/cm in the liquid) but is in agreement with EEE measured in [48] (≈0.5 at 3.0 kV/cm). In our prediction of the CEνNS effect (see below) we used EEE equal to 0.5 since we consider the dependence obtained in [48] more reliable: it was used in previous tests with our participation [51], [52], and the obtained results were self-consistent.

### 4.4 SE noise and CEνNS

The count rate of "spontaneous" SE in the test of RED-100 without passive shielding and carried out in a ground-level laboratory, i.e. without overburden from cosmic muons, is found to be ~ 250 kHz. As recently discussed in [43], there are two components in the delayed SE emission: the first one is caused by the spontaneous emission of the un-emitted electrons (namely this component is suppressed when the shutter is on), and second one which origin is probably explained by trapping and subsequent releasing the ionization electrons by the electronegative impurities. It was pointed out in [43] on possible relation of the intensity of the second component with the liquid xenon purity. Similar correlation of the rate of such delayed S2 events with purity was observed in the DarkSide50 experiment with liquid argon [53]. We have not observed the strong correlation between the SE count rate and the liquid xenon purity. Indeed, the count rate decreased during the run, while the electron lifetime increased, however in different degrees. Note, that the value of the SE count rate was obtained when the electron shutter was on. When the shutter was off the SE count rate was approximately by a factor of 3 higher. This count rate is much higher than our expectation in early publications [54], and we will continue our efforts on its reduction. Accidental coincidences of "spontaneous" SE events produce a significant background in the range of the CEνNS signal having several (≥ 2) ionization electrons (multi electron events or ME).

Note, that in this paper we consider only this, instrumental type of background which may cause principle restriction on detection of events with extremely small ionization signals. However, even with the obtained level of the SE count rate the detection of CEνNS at KNPP is feasible if a non-point-like cut is applied to the ME events. We consider true physical events as point-like ones, although they are emitted from the area of the liquid xenon surface localized within a few-mm-size and appear within ~ 0.5 μs time interval. This small non-pointness of the true ME events is caused by a transverse and longitudinal spread of an electron track in a course of electron diffusion. The cut is based on the difference in distributions of the S2 light intensity over the PMT array for point-like and non-point-like events. The following simple algorithm is used in this cut. The XY plane of S2 light sources is divided into triangles having vertexes at the PMTs centers. At the 1-st step, a triangle with the maximum sum of PMT signals is found, and at the 2-nd step, the ratio $r$ of the signal sum from this triangle to the total signal sum from the entire PMT array is obtained as a basic parameter of this cut. A Monte Carlo simulation of the light distribution for the point-like ME events and accidentally coincided SE ones was done. The best suppression factor of accidental ME was obtained at $r$ = 0.4, with the acceptance efficiency of the CEνNS events equal to 60% on average.

We estimated the expected count rate of the ME events having a specific number of electrons starting from 2 for the RED-100 detector operation at KNPP. The result is summarized in table 2. Accidental coincidence count rate was obtained from assumed Poissonian flow of the SE events with a frequency of 50 kHz. This frequency follows from 250 kHz measured in our ground-level MEPhI laboratory by applying the average reduction factor of the muon flux equal ≈5 [55]. A non-point-like cut is applied both to the accidental coincidence and CEνNS events. The CEνNS signal was estimated with the use of the latest NEST 2.01 version based on the recent measurement of ionization yield for xenon nuclear recoils in the sub-keV energy range [44]. Note, that this measurement has shown the much lower ionization yield than that predicted for the sub-keV energy range by the previous NEST 2.0 version based on the LUX data [56] and used by us in our previous publication. The CEνNS signal is normalized to 160 kg of xenon in the total sensitive volume since a position of accidental coincidence ME event cannot be defined.

Table 2. Estimated ME background at KNPP and expected CEνNS signal.

| ME value in electrons | Estimated ME background at KNPP, events/160kg/day | | Expected CEνNS count rate at KNPP, events/160kg/day | |
|---|---|---|---|---|
| | no cut | point-like | no cut | point-like |
| 2 | $5.3\cdot10^{7}$ | $1.8\cdot10^{7}$ | 465 | 283 |
| 3 | $4.4\cdot10^{5}$ | $0.9\cdot10^{5}$ | 129 | 79 |
| 4 | $2.7\cdot10^{3}$ | 348 | 35.5 | 21.7 |
| 5 | 13.7 | 1.1 | 10.6 | 6.4 |
| 6 | $5.7\cdot10^{-2}$ | $3.0\cdot10^{-3}$ | 1.9 | 1.2 |

As follows from table 2, the ME background from accidental coincidences of "spontaneous" SE becomes lower than the CEνNS signal when the events with number of electrons > 4 are selected. However, a radioactive background (n and γ) at the detector location at KNPP should be taken into account also for the final decision on the selection of a threshold level.

Our future efforts will be focused on reduction of the background caused by accidental coincidences of “spontaneous” SE count rate. Here, we plan to increase the SE signal by the increase of electric field between the anode and G2 electrode. This will result in better SPE statistics in the SE signals and better selection of point-like events. Another direction will be the reduction of “spontaneous” SE count rate. We expect that decreasing of this rate will continue during longer period of purification than that of the latest run provided these SE events are related to some specific impurities. Also, we plan to introduce selection of the quietest periods of observation (with loss of a live time, however) by implication of a smart blocking for the muon events. These events and the long tails after them will be blocked for the period of time depending on the deposited energy. For the highest values, the blocking time may reach even hundreds of milliseconds. Increasing of electric field will result in higher EEE as well. This will shift the CEνNS signal to the higher value (number of electrons in signal).

## 5. Conclusion

The first one-month long physical test of the RED-100 two-phase emission detector was carried out in February 2019 in a ground-level laboratory without overburden from cosmic muons and without passive shielding from environmental radioactivity. During this test, the following results have been obtained:

- the electron lifetime achieved the level of several milliseconds at the end of ~ 1-month period of circulation through the hot getter;
- single electron calibration and calibration by γ-sources $^{22}$Na and $^{60}$Co was performed, and the electron emission efficiency of 0.54 ± 0.08 at electric field of 3.0 ± 0.1 kV/cm was obtained which is rather different from the value obtained in [45], [46], [47] but is closer to the earlier result of [48];
- the rate of “spontaneous” SE was obtained to be ~ 250 kHz; the rate of background caused by accidental coincidences of “spontaneous” SE is estimated from this value for the experimental site at KNPP taking into account reduction of the average muon flux by a factor of 5;
- the expected CEνNS signal is recalculated with the use of the latest measurements of ionization yield of xenon nuclear recoils in the sub-keV energy range; it was demonstrated that with the current value of EEE the SE accidental coincidences background becomes lower than the CEνNS signal level when > 4 electrons are selected and detection of CEνNS signal is feasible at KNPP.

## Acknowledgments

The authors are grateful to the Russian Science Foundation for the financial support of the project under contract No.18-12-00135 from 12.04.2018, Competitiveness Improvement Program NRNU MEPhI (contract No.02.a03.21.0005, 27.08.2013), Russian Foundation for Basic Research (contracts No.18-32-00910 and 18-32-00920).